\documentclass[12pt]{article}

\usepackage{amssymb}
\usepackage{amsmath}
\usepackage{amscd}
\usepackage{latexsym}
\usepackage{graphicx}

\usepackage{caption}

\usepackage{cite}
\newcommand{\be}{\begin{equation}}
\newcommand{\ee}{\end{equation}}
\newcommand{\Dlt}{\Delta}
\newcommand{\dlt}{\delta}

\newcommand{\br}{{\bf r}}

\newcommand{\bfe}{{\bf e}}
\newcommand{\bn}{{\bf n}}
\newcommand{\bS}{{\bf S}}

\newcommand{\bB}{{\bf B}}
\newcommand{\bt}{\beta}
\newcommand{\vp}{\varphi}

\newcommand{\al}{\alpha}

\newcommand{\gm}{\gamma}
\newcommand{\om}{\omega}
\newcommand{\Om}{\Omega}

\newcommand{\rgl}{\rangle}
\newcommand{\lgl}{\langle}

\begin{document}

\begin{center}

{\Large{\bf Method of dynamic resonance tuning in spintronics of nanosystems} \\ [5mm]

V.I. Yukalov$^{1,2}$ and E.P. Yukalova$^{3}$ }  \\ [3mm]

{\it
$^1$Bogolubov Laboratory of Theoretical Physics, \\
Joint Institute for Nuclear Research, Dubna 141980, Russia \\ [2mm]

$^2$Instituto de Fisica de S\~ao Carlos, Universidade de S\~ao Paulo, \\
CP 369, S\~ao Carlos 13560-970, S\~ao Paulo, Brazil \\ [2mm]

$^3$Laboratory of Information Technologies, \\
Joint Institute for Nuclear Research, Dubna 141980, Russia } \\ [3mm]

{\bf E-mails}: {\it yukalov@theor.jinr.ru}, ~~ {\it yukalova@theor.jinr.ru}

\end{center}

\vskip 1cm

\begin{abstract}

A method is advanced allowing for fast regulation of magnetization direction 
in magnetic nanosystems. The examples of such systems are polarized nanostructures, 
magnetic nanomolecules, magnetic nanoclusters, magnetic graphene, dipolar and spinor 
trapped atoms, and quantum dots. The emphasis in the paper is on magnetic nanomolecules 
and nanoclusters. The method is based on two principal contrivances: First, the magnetic 
sample is placed inside a coil of a resonant electric circuit creating a feedback 
field, and second, there is an external magnetic field that can be varied so that to 
dynamically support the resonance between the Zeeman frequency of the sample and 
the natural frequency of the circuit during the motion of the sample magnetization. 
This method can find applications in the production of memory devices and other 
spintronic appliances.       

\end{abstract}

\vskip 3mm
 
\newpage

\section{Introduction}

In the problem of regulating spin dynamics, there are two principal challenges 
contradicting each other. First, it is necessary to possess the ability of keeping 
fixed the device magnetization for sufficiently long time. And second, one has to 
have the capability of quickly varying the magnetization direction at any required 
time. The possibility of keeping fixed the magnetization direction, as such, is not 
a difficult task that can be easily realized by using the materials enjoying strong 
magnetic anisotropy. The well known examples of materials with strong magnetic 
anisotropy are magnetic nanomolecules 
\cite{Barbara_1,Caneschi_2,Yukalov_31,Yukalov_3,Yukalov_32,Yukalov_33,Friedman_4,
Miller_5,Craig_6,Liddle_7,Rana_8} and magnetic nanoclusters 
\cite{Kodama_9,Hadjipanays_10,Wernsdorfer_11,Yukalov_34,Kudr_12}.
These materials can possess large spins and strong magnetic anisotropy keeping, at 
low temperature, below the blocking temperature, the sample magnetization frozen. 
At the same time, the strong magnetic anisotropy prevents the realization of fast 
magnetization reversal because of two reasons. First, as will be shown below, 
magnetic anisotropy induces the dynamic variation of the effective Zeeman frequency 
which cannot be compensated by a constant magnetic field, and second, even inverting 
the external magnetic field, the sample spin cannot be quickly reversed because of 
the strong magnetic anisotropy. Thus the dilemma arises: For fixing during the 
required long time the magnetization direction, one needs a rather strong magnetic 
anisotropy; however the latter does not allow for fast magnetization reversal. 

In the present paper, we advance an original way out of the above dilemma. We 
keep in mind the materials enjoying magnetic anisotropy sufficient for fixing the 
sample magnetization. These can be magnetic nanomolecules or magnetic nanoclusters. 
To some extent, the consideration is applicable to dipolar and spinor trapped atoms 
\cite{Griesmaier_38,Baranov_39,Baranov_40,Stamper_41,Gadway_42,Yukalov_35,Yukalov_43} 
and to magnetic graphene (graphene with magnetic defects) \cite{Yukalov_36,Yukalov_37}. 
Quantum dots in many aspects are similar to nanomolecules \cite{Birman_13} and also 
can possess magnetization \cite{Schwartz_14,Koole_15,Mahagan_16,Tufany_17} that could 
be governed.      

The method we suggest is based on two principal points. First, the considered 
sample is placed inside a magnetic coil of an electric circuit. Then the moving 
magnetization of the sample produces electric current in the circuit, which, in 
turn, creates a magnetic feedback field acting on the sample. An effective coupling 
between the circuit and the sample appears only when the circuit natural frequency 
is in resonance with the Zeeman frequency of the sample. However, the magnetic 
anisotropy, as we show, leads to the detuning of the effective Zeeman frequency 
from the resonance. Moreover, this detuning is dynamic, varying in time. The second 
pillar of the suggested method is the use of a varying external magnetic field 
realizing the dynamic tuning of the effective Zeeman frequency to the resonance with 
the circuit natural frequency. Since the tuning procedure is dynamic, the method is 
called the {\it dynamic resonance tuning}.

\section{Magnetic nanomolecules and nanoclustres}

Single-domain nanomolecules and nanoclusters are of special interest for spintronics. 
Due to strong exchange interactions, the spins of particles forming the nanomagnet 
point in the same direction thus creating a common spin and hence the common single
magnetization vector. Under the action of external fields, the magnetization vector
moves as a whole, which is called coherent motion. At the same time, such nanomagnets
usually possess a rather strong magnetic anisotropy. The standard Hamiltonian of a 
nanomagnet with a spin $S$ has the form
\be
\label{1}
\hat H = -\mu_S \bB \cdot \bS + \hat H_A \; ,
\ee
where the first is the Zeeman term, $\mu_S = - g_S \mu_B$, $g_S$ is a Land\'e 
factor, $\mu_B$ is the Bohr magneton, and ${\bf S}$ is a spin operator. The second 
term describes the nanomagnet magnetic anisotropy,
\be
\label{2}
 \hat H_A = - D S_z^2 + E ( S_x^2 - S_y^2 ) \; ,
\ee
with the anisotropy parameters
$$
D = \frac{1}{2} \; ( D_{xx} + D_{yy} ) - D_{zz} \; , 
\qquad
E = \frac{1}{2} \; ( D_{xx} - D_{yy} )  \; ,   
$$
expressed through the dipolar tensor $D_{\alpha \beta}$ and $S_\alpha$ being 
spin-operator components. 

The sample is inserted into a magnetic coil of an electric circuit producing a 
feedback field $H$. The coil axis is in the $x$ direction. The total magnetic 
field, acting on the sample,
\be
\label{3}
  \bB = H \bfe_x +B_1 \bfe_y + ( B_0 + \Dlt B ) \bfe_z \; ,
\ee
consists of the feedback field $H$, a small anisotropy field $B_1$, and an external 
magnetic field $B_0+\Dlt B$, with $B_0$ being a constant field and $\Dlt B$ being 
a regulated part of the field. The electric circuit plays the role of a resonator. 
The motion of the sample magnetization creates in the circuit electric current 
satisfying the Kirhhoff equation. In its turn, the electric current forms the 
feedback magnetic field defined by the equation \cite{Yukalov_18} following from
the Kirhhoff equation,
\be
\label{4}
 \frac{dH}{dt} + 2\gm H + \om^2 \int_0^t H(t') \; dt' = - 
4\pi\eta_{res} \; \frac{d m_x}{dt} \;  ,
\ee
where $\gm$ is the circuit attenuation, $\om$, the resonator natural frequency,
and $\eta_{res}\approx V/V_{res}$ is the resonator coil filling factor, $V$ being 
the sample volume, and $V_{res}$, the resonator coil volume. The right-hand side of
the equation characterizes the electromotive force due to the moving average 
magnetization
\be
\label{5}
 m_x = \frac{\mu_S}{V} \; \lgl \; S_x \;\rgl \;  .
\ee
       
Writing down the Heisenberg equations of motion, we average them, looking for the 
dynamics of the average spin components
\be
\label{6}
x = \frac{\lgl \; S_x\; \rgl}{S} \;  , 
\qquad
y = \frac{\lgl \; S_y\; \rgl}{S} \;  , 
\qquad
z = \frac{\lgl \; S_z\; \rgl}{S} \;  , 
\ee
with $S$ being the sample spin. In the process of the averaging, we meet the 
combination of spins $S_\al S_\bt + S_\bt S_\al$. It would be incorrect to 
decouple this combination in the simple mean-field approximation, since for 
$S=1/2$ this combination has to be zero. The correct decoupling \cite{Yukalov_19}, 
that is exactly valid for $S=1/2$ as well as asymptotically exact for large spins, 
is done in the corrected mean-field approximation
\be
\label{7}
 \lgl \; S_\al S_\bt+ S_\bt S_\al \; \rgl = \left( 2 - \; \frac{1}{S} \right) \;
\lgl \; S_\al \; \rgl \lgl \; S_\bt \; \rgl \; .
\ee

For what follows, we need to introduce several notations. We define the Zeeman frequency
\be
\label{8}
 \om_0 \equiv -\; \frac{\mu_S}{\hbar} \; B_0 \;  ,   
\ee
and the anisotropy frequencies
\be
\label{9}
 \om_1 \equiv -\; \frac{\mu_S}{\hbar} \; B_1 \;  ,  
\qquad
\om_D \equiv (2S - 1) \; \frac{D}{\hbar}  \;  ,
\qquad
\om_E \equiv (2S - 1) \; \frac{E}{\hbar}  \;  .
\ee
The dimensionless anisotropy parameter is defined as
\be
\label{10}
 A \equiv  \frac{\om_D+\om_E}{\om_0}  \;  .
\ee
The dimensionless regulated field is given by the expression
\be
\label{11}
 b \equiv -\; \frac{\mu_S\Dlt B}{\hbar\om_0}  \;   .
\ee

From the evolution equations, it is seen that the coupling between the resonant 
circuit and the sample is characterized by the coupling rate
\be
\label{12}
 \gm_0 \equiv \pi \eta_{res} \; \frac{\mu_S^2 S}{\hbar V} =
\pi \; \frac{\mu_S^2 S}{\hbar V_{res}}  \;   .
\ee
Finally, the dimensionless feedback field is denoted by
\be
\label{13}
 h \equiv -\;\frac{\mu_S H}{\hbar \gm_0} \;  .
\ee

Thus we come to the system of equations
$$
\frac{dx}{dt} = - \om_S y + \om_1 z \; , \qquad
\frac{dy}{dt} =  \om_S x - \gm_0 h z \; ,
$$
\be
\label{14}
 \frac{dz}{dt} = 2 \om_E x y - \om_1 x + \gm_0 h y \; ,
\ee
in which the effective Zeeman frequency is
\be
\label{15}
 \om_S \equiv \om_0 \; ( 1 + b - A z ) \; .
\ee
The feedback-field equation (\ref{4}) becomes
\be
\label{16}
 \frac{dh}{dt} + 2\gm h + \om^2 \int_0^t h(t') \; dt' =  4\; \frac{dx}{dt} \;  .
\ee   

Of course, the evolution equations are to be complemented by the initial conditions
${\bf x}(0) = \bf{x}_0$ for the vector ${\bf x} = \{x,y,z\}$ and for the feedback 
field $h(0) = h_0$. In what follows, we set the initial conditions as $x_0=y_0=0$,
$h_0=0$, and $z_0=1$. These initial conditions correspond to the setup, when there
are no alternating fields pushing the spin motion and all the following dynamics 
is self-organized.

\section{Dynamic resonance tuning for single spins}

The initial setup $s_0=1$ describes the sample with the spin polarization, formed 
by electrons, directed up, hence the magnetization directed down, while the external 
magnetic field $B_0$ is directed up. This implies that the sample is in a metastable 
state. The stable state corresponds to the magnetization up, hence to the spin 
polarization down. 

Because of the strong magnetic anisotropy, the magnetization is frozen and, below 
the blocking temperature, it can exist in this metastable state for days and months. 
This is convenient for keeping the information in memory devices. However, if on 
needs to rewrite or erase the information, which also is an action typical of memory 
devices, then the frozen magnetization hinders this. Thus we confront the problem: 
how could we overcome the anisotropy in order to start moving the sample spin and 
to move it sufficiently fast? 

If there would be resonance between the resonator natural frequency $\om$ and the 
effective Zeeman frequency $\om_S$, which is possible in the absence of the magnetic 
anisotropy, when $A=0$, then there would appear strong coupling between the resonator
feedback field and the sample magnetization, as a result of which the magnetization 
could be quickly reversed at the initial time 
\cite{Yukalov_18,Yukalov_19,Yukalov_20,Yukalov_21,Kharebov_22}. However, the effective 
Zeeman frequency (\ref{15}) cannot be tuned to a constant resonator natural frequency 
$\om$, since the effective Zeeman frequency is not a constant but a function of the 
polarization $z$. For instance, if $\om$ is tuned to $\om_0$, then the relative 
detuning
$$
\frac{\om_S -\om_0}{\om_0} = b - Az
$$ 
varies in time and, generally, can be very large for large anisotropy parameters $A$.

Suppose, we have been keeping the magnetization fixed for a required time $\tau$, 
after which we need to quickly reverse it. To make the detuning small, and moreover 
for keeping it small during the whole process of spin reversal, we suggest to set 
$\om_0=\om$ and to vary the regulated field $b=b(t)$ according to the law of {\it 
dynamic resonance tuning}, so that  
\begin{eqnarray}
\label{17}
b(t) = \left\{ \begin{array}{ll}
0 \; ,        ~ & ~ t < \tau \\
A z_{reg}\; , ~ & ~ t \geq \tau \; ,
\end{array} \right.
\end{eqnarray}
with $z_{reg}$ satisfying the equations
$$
\frac{dx_{reg}}{dt} = - \om_0 y_{reg} + \om_1 z_{reg} \; , \qquad
\frac{dy_{reg}}{dt} =  \om_0 x_{reg} - \gm_0 h_{reg} z_{resg} \; ,
$$
$$
\frac{dz_{reg}}{dt} = 
2 \om_E x_{reg} y_{reg} - \om_1 x_{reg} + \gm_0 h_{reg} y_{reg} \; ,
$$
\be
\label{18}
\frac{dh_{reg}}{dt} + 2\gm h_{reg} + \om^2 \int_\tau^t h_{reg}(t') \; dt' =  
4\; \frac{dx_{reg}}{dt} \;  .
\ee
The initial conditions for the latter system of equations can be taken either 
as ${\bf x}_{reg}(\tau)={\bf x}(\tau)$ and $h_{reg}(\tau)=h(\tau)$ or as           
${\bf x}_{reg}(\tau)={\bf x}_0$ and $h_{reg}(\tau)=h_0$.   

In Fig. 1, we show the process where the spin polarization is frozen, by a strong 
magnetic anisotropy, during the time $\tau$, after which the mechanism of the 
dynamic resonance tuning is switched on. Different initial conditions are compared, 
as explained in the Figure, demonstrating that they lead to a slight shift of the 
reversal process. It is also shown that if the resonance condition is not dynamic, 
but only the initial triggering resonance \cite{Yukalov_23}, when the condition 
$b(\tau) = A z(\tau)$ is imposed, then  the reversal is not fast but possesses 
a long tail. Time is measured in units of $\gamma_0^{-1}$ and frequencies, in units 
of $\gamma_0$.   

%Figure 1
\begin{figure}[ht]
\centerline{
\hbox{ \includegraphics[width=7cm]{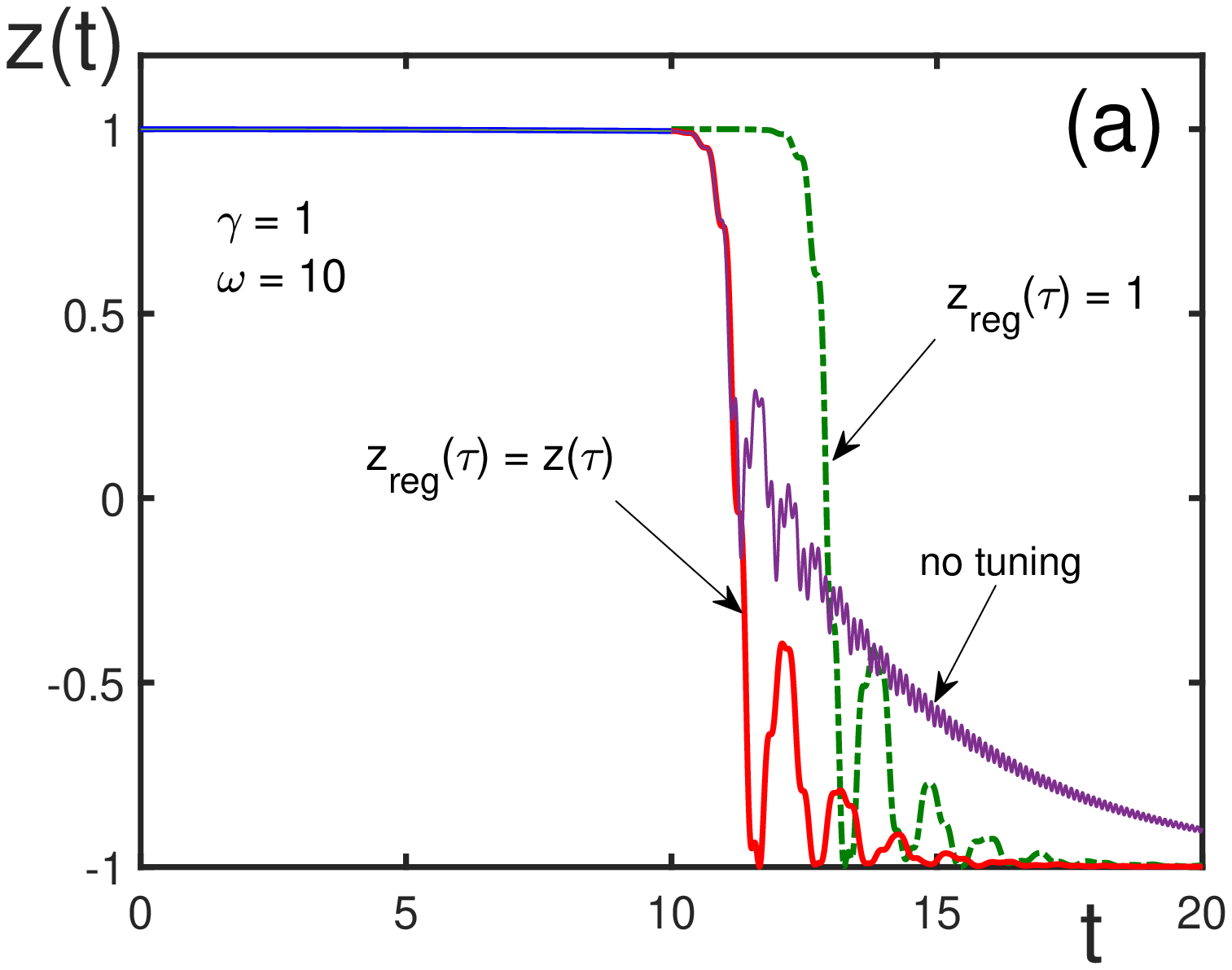} \hspace{0.5cm}
\includegraphics[width=7cm]{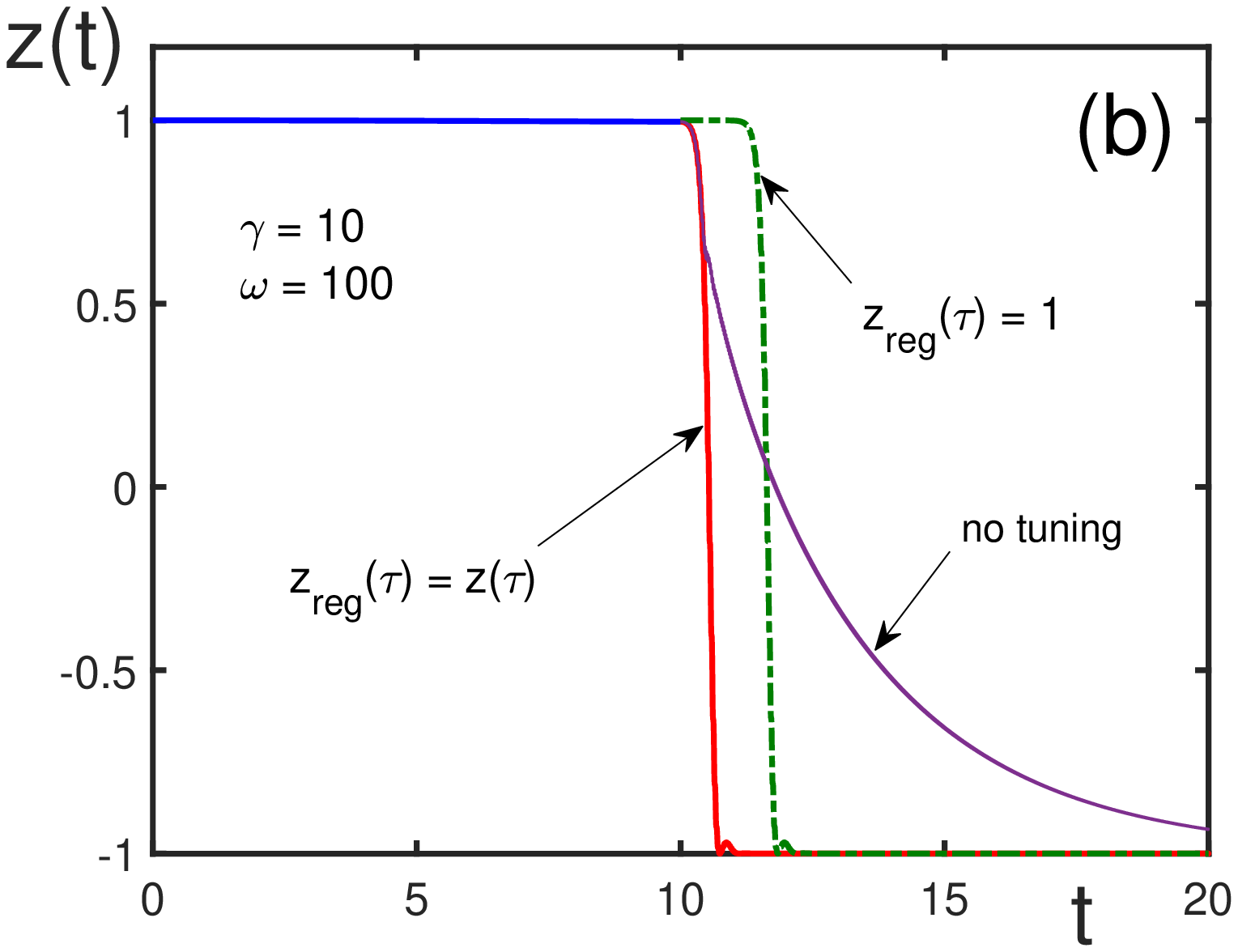}  } }
\caption{\small
Spin polarization $z(t)$ of a nanocluster or nanomolecule as a function of 
time under dynamic resonance tuning, starting at the delay time $\tau=10$, 
for different initial conditions, as explained in the Figure. The anisotropy 
parameters are $A=1$ and $\om_E=\om_1=0.01$. The line with a long tail describes 
the process, where the dynamic resonance tuning is not used, but instead the 
condition of triggering resonance at the same initial time is employed. 
(a) $\om=\om_0=10$ and $\gm=1$; (b) $\om=\om_0=100$ and $\gm=10$. Time is 
measured in units of $\gm_0^{-1}$ and frequencies, in units of $\gm_0$.  
}
\label{fig:Fig.1}
\end{figure}

In Fig. 2, the spin polarization of a nanocluster or nanomolecule is shown for
dynamic resonance tuning, compared with the polarization without dynamic tuning 
but under the triggering resonance at the same delay time. The advantage of dynamic 
resonance tuning is in an ultrafast spin reversal, while the triggering resonance
leads to long tails. 

%Figure 2
\begin{figure}[ht]
\centerline{
\hbox{ \includegraphics[width=7cm]{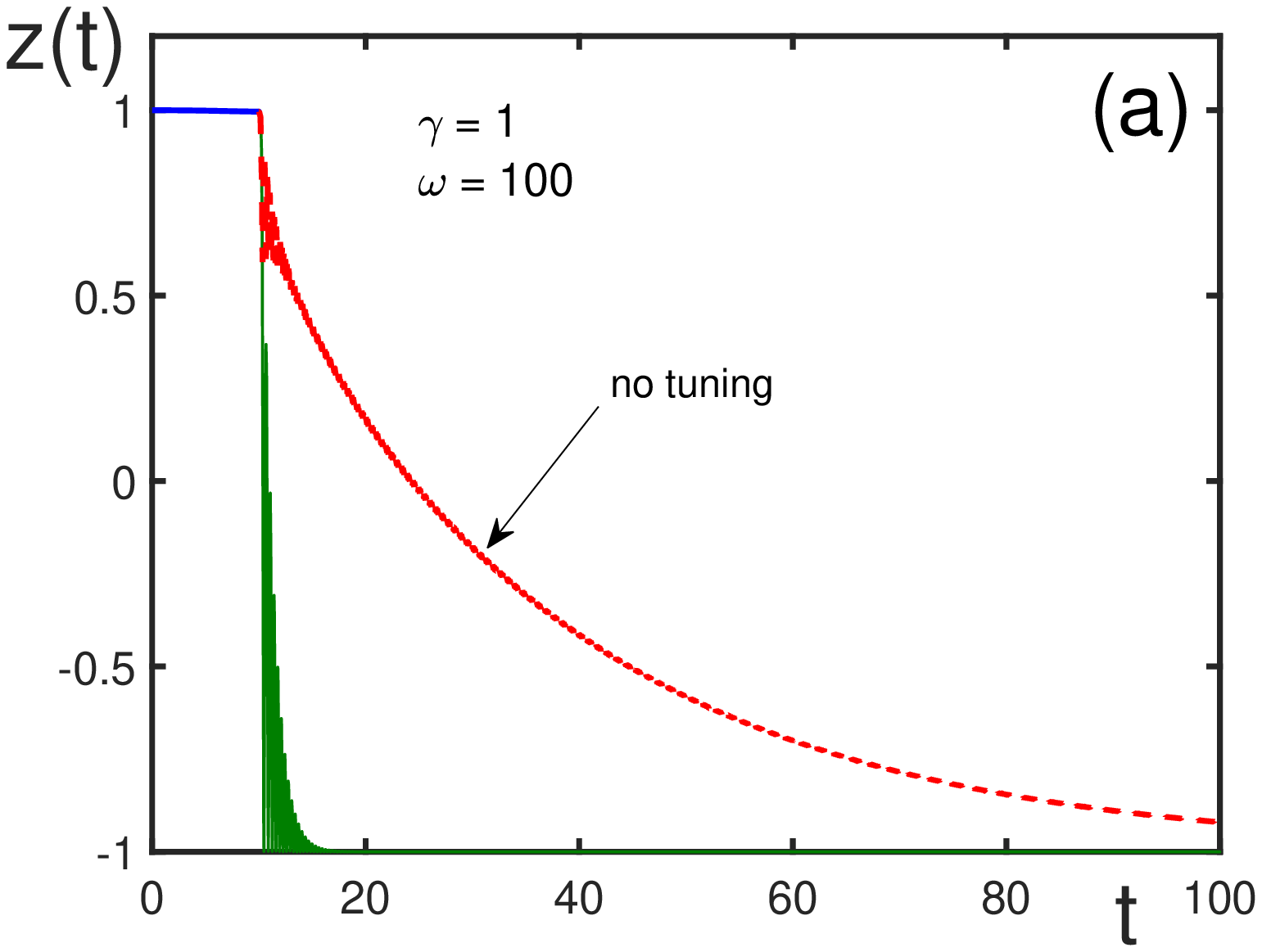} \hspace{0.5cm}
\includegraphics[width=7cm]{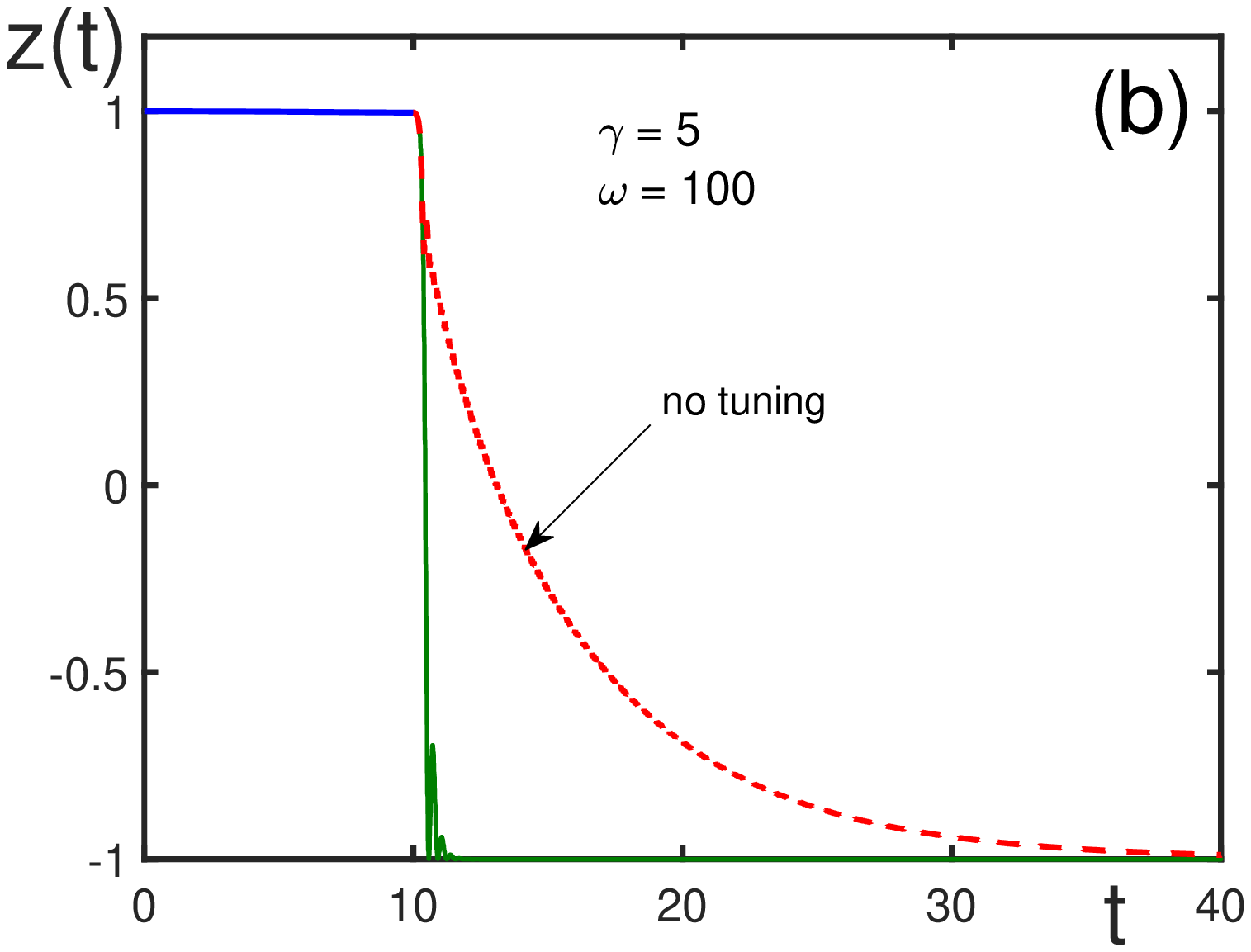}  } }
\caption{\small 
Spin polarization $z(t)$ of a nanocluster or nanomolecule as a function of time 
under dynamic resonance tuning, starting at the delay time $\tau=10$ (solid line), 
compared with the polarization without dynamic tuning but under the triggering 
resonance at the same delay time (dashed line). The anisotropy parameters are 
$A=1$ and $\om_E=\om_1=0.01$. 
(a) $\om=\om_0=100$ and $\gm=1$; 
(b) $\om=\om_0= 100$ and $\gm=5$. The absence of dynamic tuning leads to long 
tails. 
}
\label{fig:Fig.2}
\end{figure}

Figure 3 illustrates the reversal of the spin polarization of a nanocluster or 
nanomolecule under dynamic resonance tuning, starting at different delay times. This
shows that it is possible to quickly reverse the magnetization at any required time. 

%Figure 3
\begin{figure}[ht]
\centerline{
\hbox{ \includegraphics[width=8cm]{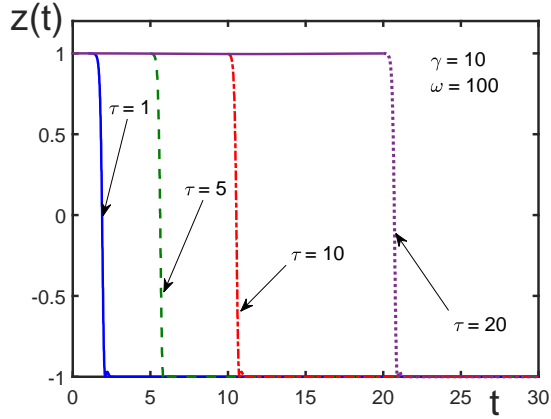}  } }
\caption{\small
Spin polarization $z(t)$ of a nanocluster or nanomolecule as a function of time 
under dynamic resonance tuning, starting at different delay times: $\tau=1$ (solid 
line), $\tau=5$ (dashed line), $\tau=10$ (dash-dotted line), and $\tau=20$ (dotted 
line). Other parameters are fixed as $A=1$, $\om_E=\om_1=0.01$, $\om=\om_0=100$ 
and $\gm=10$.
}
\label{fig:Fig.3}
\end{figure}

\section{Assemblies of nanomolecules or nanoclusters}

In the previous section, we have considered the realization of dynamic resonance 
tuning for single nanomolecules or nanoclusters, which allows for fast magnetization 
reversal at any required time. The natural question is whether it would be feasible 
to apply this effect for regulating spin dynamics of the assemblies of nanomolecules 
or nanoclusters.

The Hamiltonian of a system of nanomolecules or nanoclusters reads as
\be
\label{19}
 \hat H = -\mu_S \sum_j \bB \cdot \bS_j + \hat H_A +\hat H_D \;  ,
\ee
where $j = 1,2, \ldots, N$ enumerates the clusters, the first is the Zeeman term 
and the second is the anisotropy term
\be
\label{20}
\hat H_A = - \sum_j D(S_j^z)^2 \; .
\ee
The anisotropy parameter $E$ is usually much smaller than  $D$, so it can be safely 
omitted. 

In addition, there is the term responsible for the dipolar interactions of the 
constituents
\be
\label{21}
 \hat H_D = \frac{1}{2} \sum_{i\neq j} \;
\sum_{\al\bt} D_{ij}^{\al\bt} S_i^\al S_j^\bt \;  ,
\ee
with the dipolar tensor
\be
\label{22}
D_{ij}^{\al\bt} = \frac{\mu_S^2}{r_{ij}^3} \;\left( \dlt_{\al\bt} - 
 3 n_{ij}^\al n_{ij}^\bt\right) \;  ,
\ee
in which
$$
r_{ij} \equiv |\; \br_{ij} \; | \; , \qquad
\bn_{ij} \equiv \frac{\br_{ij}}{r_{ij}} \; , \qquad
\br_{ij} \equiv \br_i - \br_j \; .
$$
  
The sample is again inserted into a magnetic coil of an electric circuit. The total 
magnetic field is the sum
\be
\label{23}
\bB = H \bfe_x + ( B_0 + \Dlt B ) \bfe_z
\ee
of the feedback field $H$, a constant external magnetic field $B_0$, and a 
regulated field $\Delta B$.  

The feedback field satisfies equation (\ref{4}), but with the right-hand side, 
describing the electromotive force, containing the average magnetization 
\be
\label{24}
 m_x = \frac{\mu_S}{V} \sum_j \lgl \; S_j^x \;\rgl \;  .
\ee
The coil axis is again aligned with the $x$ axis. 

In what follows, it is convenient to employ the ladder spin operators 
$S_j^{\pm}=S_j^x\pm i S_j^y$. We introduce the average transverse spin component
\be
\label{25}
u = \frac{1}{SN} \sum_j \lgl \; S_j^- \;\rgl \; ,
\ee
the coherence intensity
\be
\label{26}
 w = \frac{1}{SN(N-1)} \sum_{i\neq j} \lgl \; S_i^+ S_j^- \;\rgl \;  ,
\ee
and the longitudinal spin polarization
\be
\label{27}
s = \frac{1}{SN} \sum_j \lgl \; S_j^z \;\rgl \;   .
\ee

Local spin fluctuations, caused by dipolar interactions, are characterized by the 
expressions
$$
\xi_S = \frac{1}{\hbar} \sum_j \lgl \; a_{ij} S_j^z  + c_{ij} S_j^+
+ c_{ij}^* S_j^- \;\rgl \; ,
$$
\be
\label{28}
\vp_S = \frac{1}{\hbar} \sum_j \left\lgl \; \frac{a_{ij}}{2}\; S_j^-
  - 2b_{ij} S_j^+ - 2c_{ij} S_j^z \; \right\rgl \; ,
\ee
in which
$$
a_{ij} \equiv D_{ij}^{zz} \; , \qquad 
b_{ij} \equiv \frac{1}{4} \;\left( D_{ij}^{xx} - D_{ij}^{yy} - 
2i D_{ij}^{xy} \right) \; ,
\qquad
c_{ij} \equiv \frac{1}{2} \;\left( D_{ij}^{xz} - i D_{ij}^{yz} \right) \; .
$$
These expressions are responsible for dipolar spin waves initiating spin motion at
the beginning of the process \cite{Yukalov_19}. 

The decoupling of pair spin correlators is accomplished in the corrected mean-field
approximation 
$$
\lgl \; S_i^\al S_j^\bt \; \rgl = 
\lgl \; S_i^\al \; \rgl \lgl \; S_j^\al \;\rgl \qquad ( i \neq j) \; ,
$$
\be
\label{29}
\lgl \; S_j^\al S_j^\bt + S_j^\bt S_j^\al \; \rgl =  
\left( 2 - \; \frac{1}{S} \right)  \; 
\lgl \; S_j^\al \; \rgl \lgl \; S_j^\bt \; \rgl
\ee
accurately taking into account the terms describing magnetic anisotropy 
\cite{Yukalov_18}. Writing down the spin equations of motion and averaging them, we 
come to the system of equations
$$
\frac{du}{dt} = - i ( \om_S + \xi_S - i\gm_2 ) \; u + f s \; ,
\qquad 
\frac{dw}{dt} = - 2 \gm_2 w  + ( u^* f + f^* u )\; s \; ,
$$
\be
\label{30}
 \frac{ds}{dt} = - \; \frac{1}{2}\;  ( u^* f + f^* u ) \;  ,
\ee
with the transverse attenuation
\be
\label{31}
 \gm_2 = \frac{1}{\hbar}\; \rho \; \mu_S^2 S \;  ,
\ee
where $\rho = N/V$ is spin density, and the effective force
\be
\label{32}
f = - i \left( \frac{\mu_S H}{\hbar} + \vp_S \right) \;  .
\ee
The effective Zeeman frequency here is
\be
\label{33}
\om_S = \om_0 ( 1 + b - As) \; ,
\ee
where the dimensionless regulated field $b$ is defined in Eq. (\ref{11}).

The feedback-field equation (\ref{4}) can be represented in the integral form
\be
\label{34}
 H = - 4\pi \eta_{res} \int_0^t G(t-t') \; \dot{m}_x(t') \; dt' \; ,
\ee
with the transfer function 
$$
G(t) = \left[\; \cos(\om' t) \; - \; 
\frac{\gm}{\om'}\;\sin(\om' t) \; \right] \; e^{-\gm t}\; ,
$$
where the frequency is
$$
 \om' \equiv \sqrt{\om^2 + \gm^2} \;  .
$$
The electromotive force is expressed through
\be
\label{35}
 \dot{m}_x = \frac{1}{2} \; \rho \; \mu_S S \; ( u^* + u ) \;  .
\ee

The coupling rate is
\be
\label{36}
\gm_0 = \pi \eta_{res} \gm_2 \; .
\ee
As usual, the attenuations are small as compared to the related frequencies:
\be
\label{37}
  \frac{\gm_0}{\om} \ll 1 \; , \qquad \frac{\gm}{\om} \ll 1 \; , \qquad 
  \frac{\gm_2}{\om_0} \ll 1 \; .
\ee

The solution to the feedback field, following from Eq. (\ref{34}), to the first 
order in $\gamma_0$ reads as
\be
\label{38}
 \mu_S H = i \hbar \; ( u X - X^* u^* ) \;  ,
\ee
with the coupling function
\be
\label{39}
 X = \gm_0 \; \om_S \; 
\frac{1 -\exp(-i\Dlt_S t - \gm t)}{\gm + i\Dlt_S \; {\rm sign}\;\om_S}  
\ee
and the dynamic detuning
\be
\label{40}
 \Dlt_S = \om - |\; \om_S \; | \;  .
\ee
   
The presence of the small parameters (\ref{37}) makes it straightforward to resort
to the averaging techniques \cite{Bogolubov_24,Freidlin_25}, with considering the
dipolar spin fluctuations, characterized by expressions (\ref{28}), as small random 
variables \cite{Gardiner_26,Yukalov_27}. Due to the small parameters (\ref{37}), 
the function $u=u(t)$ in Eqs. (\ref{30}) has to be treated as a fast variable, while 
the functions $w = w(t)$ and $s = s(t)$, as slow variables. With the slow variables 
playing the role of adiabatic invariants, we solve the equation for the fast variable, 
yielding
$$
u = u_0 \exp\left\{ - i \Om t - i \int_0^t \xi_S(t') \; dt' \right\} \; -
$$
\be
\label{41}
- \;
i s \int_0^t \vp_S(t') \; 
\exp\left\{ - i \Om(t-t') - i \int_{t'}^t \xi_S(t'') \; dt'' \right\} \; dt' \; ,
\ee
where
$$
 \Om = \om_S - i (\gm_2 - Xs ) \;  .
$$
Then, substituting the solutions for the feedback field (\ref{39}) and for the fast 
variable (\ref{41}) into the equations for the slow variables $w$ and $s$, we average
the latter over time and random spin fluctuations \cite{Yukalov_28}. As a result, we 
obtain the guiding-center equations
\be
\label{42}
 \frac{dw}{dt} = 2\gm_2 w ( \al s - 1) + 2\gm_3 s^2 \; , \qquad
\frac{ds}{dt} = -\gm_2 \al w - \gm_3 s \;  ,
\ee
in which we define the coupling function 
\be
\label{43}
\al \equiv \frac{{\rm Re} X}{\gm_2} = 
\frac{\gm_0 \gm \om_S}{\gm_2(\gm^2+\Dlt_S^2)} 
\left\{ 1 - [\; \cos(\Dlt_S t) - 
\dlt_S \sin(\Dlt_S t) \; ]\; e^{-\gm t}\right\} \; ,
\ee
the relative detuning
\be
\label{44}
\dlt_S \equiv \frac{\Dlt_S}{\gm}\; {\rm sign}\; \om_S \;   ,
\ee
and the spin-wave attenuation
\be
\label{45}
 \gm_3 = \frac{\gm_2^2}{\sqrt{\om_S^2+\gm_2^2} } \;  .
\ee
As initial conditions, we take $w(0) = 0$, which implies the absence of triggering 
fields, so that the process is self-organized, and we assume the initial average spin 
polarization up, so that $s(0) = 1$.

\section{Dynamic resonance tuning for spin assemblies}

Suppose we wish to keep the average spin polarization up until the time $\tau$ 
and then we need to quickly reverse the average spin. For this purpose, we set 
$\om_0=\om$ and arrange the regulated magnetic field according to the law
\begin{eqnarray}
\label{46}
b(t) = \left\{ \begin{array}{ll}
0 \; ,         ~ & ~ t < \tau \\
A s_{reg} \; , ~ & ~ t \geq \tau \; ,
\end{array} \right.
\end{eqnarray}
with the parameter 
\be
\label{47}
A \equiv \frac{\om_D}{\om_0} 
\ee
and with $s_{reg}$ satisfying the equations
$$
\frac{d s_{reg}}{dt}= -\gm_2 \al_{reg} w_{reg} - \gm_{reg} s_{reg} \; ,
$$
\be
\label{48}
\frac{d w_{reg}}{dt}= 2\gm_2 w_{reg} (\al_{reg} s_{reg} -1 )
+ 2 \gm_{reg} s_{reg}^2  \; .
\ee
In the latter, the effective coupling function is
\be
\label{49}
 \al_{reg} = g \left( 1 - e^{-\gm t} \right) \;  ,
\ee
with the coupling parameter
\be
\label{50}
g \equiv \frac{\gm_0 \om_0}{\gm \gm_2} 
\ee
and the attenuation
\be
\label{51}
 \gm_{reg} \equiv \frac{\gm_2^2}{\sqrt{\om^2+ \gm_2^2} } \;  .
\ee
The initial conditions at time $\tau$ are $w_{reg}(\tau)=w(\tau)$ and 
$s_{reg}(\tau)=s(\tau)$. At time $\tau$, there is the resonance $\om_0=\om$, and 
in the following times the regulated field (\ref{46}) varies so that the system is 
dynamically captured into resonance \cite{Lochak_29}. 

If there is need for repeating the overall process again, this can be done by either 
inversing the direction of the external field $B_0$ or by rotating the sample as has 
been described for organizing a Morse-code alphabet functioning of spin pulses for 
samples having no magnetic anisotropy \cite{Yukalov_30}.
    
Figures 4 and 5 demonstrate the effect of dynamic resonance tuning for the assemblies 
of magnetic nanomolecules or nanoclusters. The average spin of the system can be kept 
for a long time by a strong magnetic anisotropy. The spin is better frozen for larger
anisotropies and larger Zeeman frequencies. Employing dynamic resonance tuning makes 
it possible to realize an ultrafast spin reversal at any required time.   

%Figure 4
\begin{figure}[ht]
\centerline{
\hbox{ \includegraphics[width=7cm]{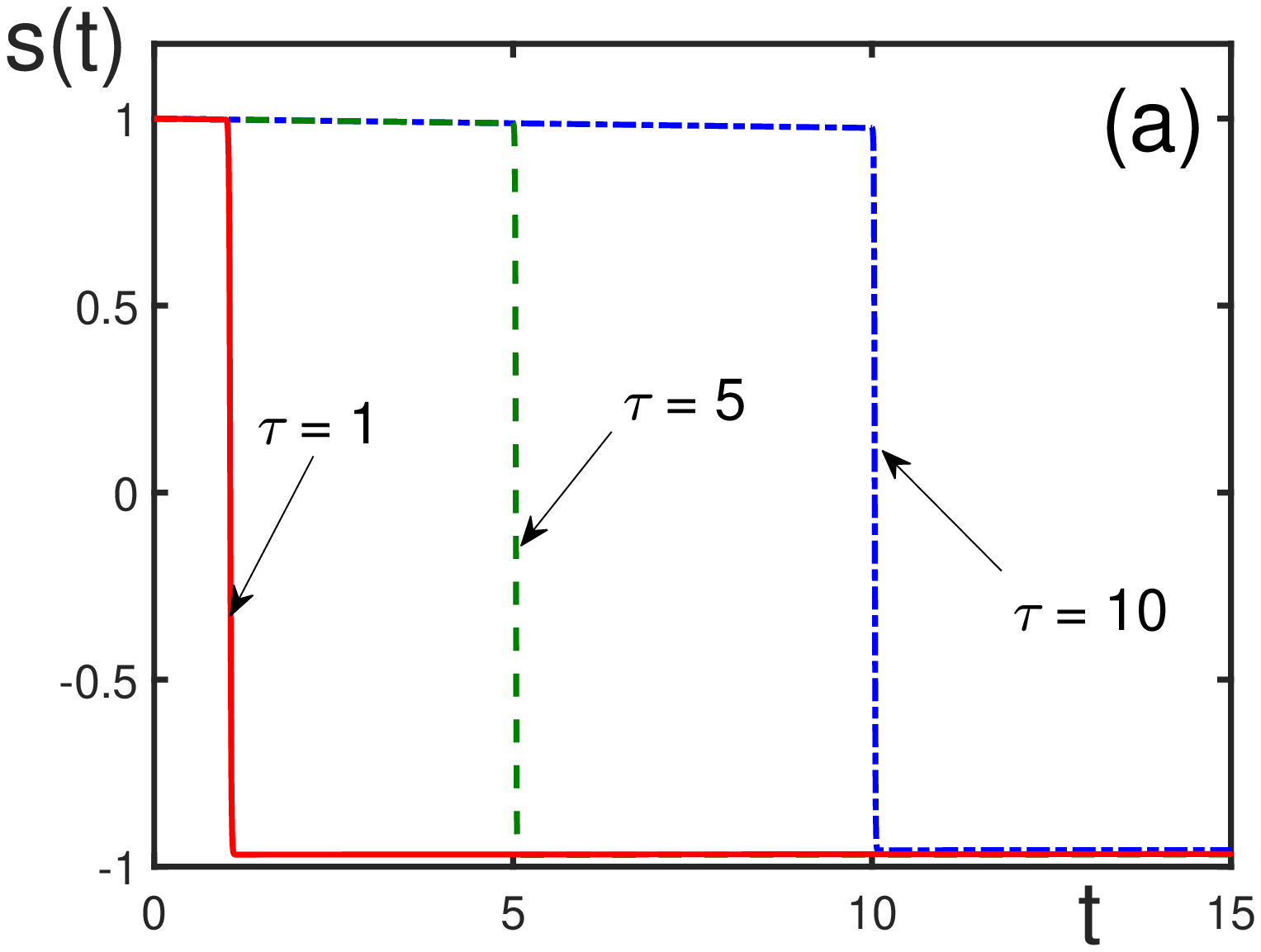} \hspace{0.5cm}
\includegraphics[width=7cm]{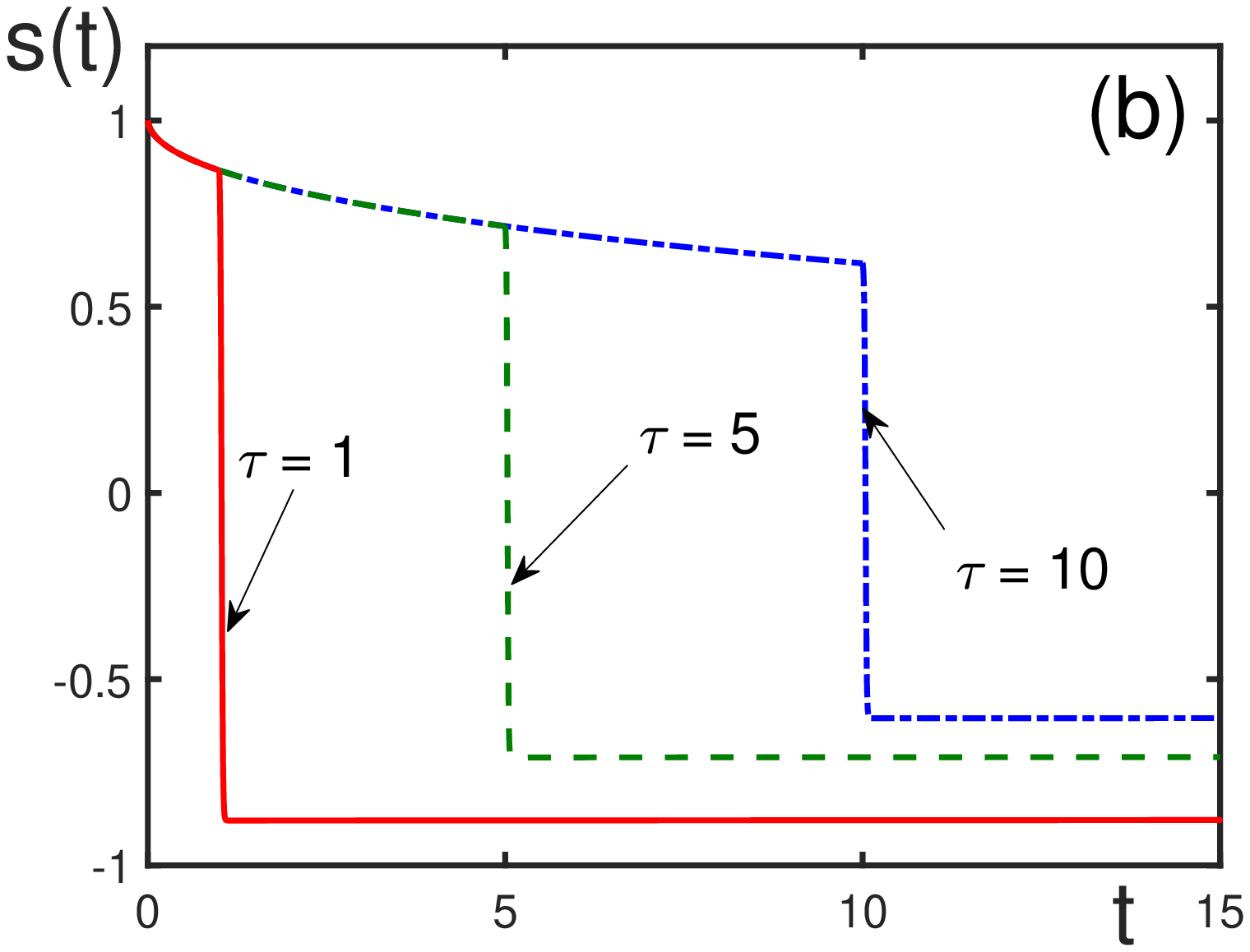}  } }
\caption{\small 
Average spin polarization $s(t)$ of an assembly of magnetic nanomolecules or 
nanoclusters for $\om=\om_0=100$ and $\gm=10$ as a function of time under dynamic 
resonance tuning starting at different delay times: $\tau=1$ (solid line), $\tau=5$ 
(dashed line), and $\tau=10$ (dash-dotted line). The anisotropy parameters are: 
(a) $A=5$; (b) $A=1$.
}
\label{fig:Fig.4}
\end{figure}

%Figure 5
\begin{figure}[ht]
\centerline{
\hbox{ \includegraphics[width=7.0cm]{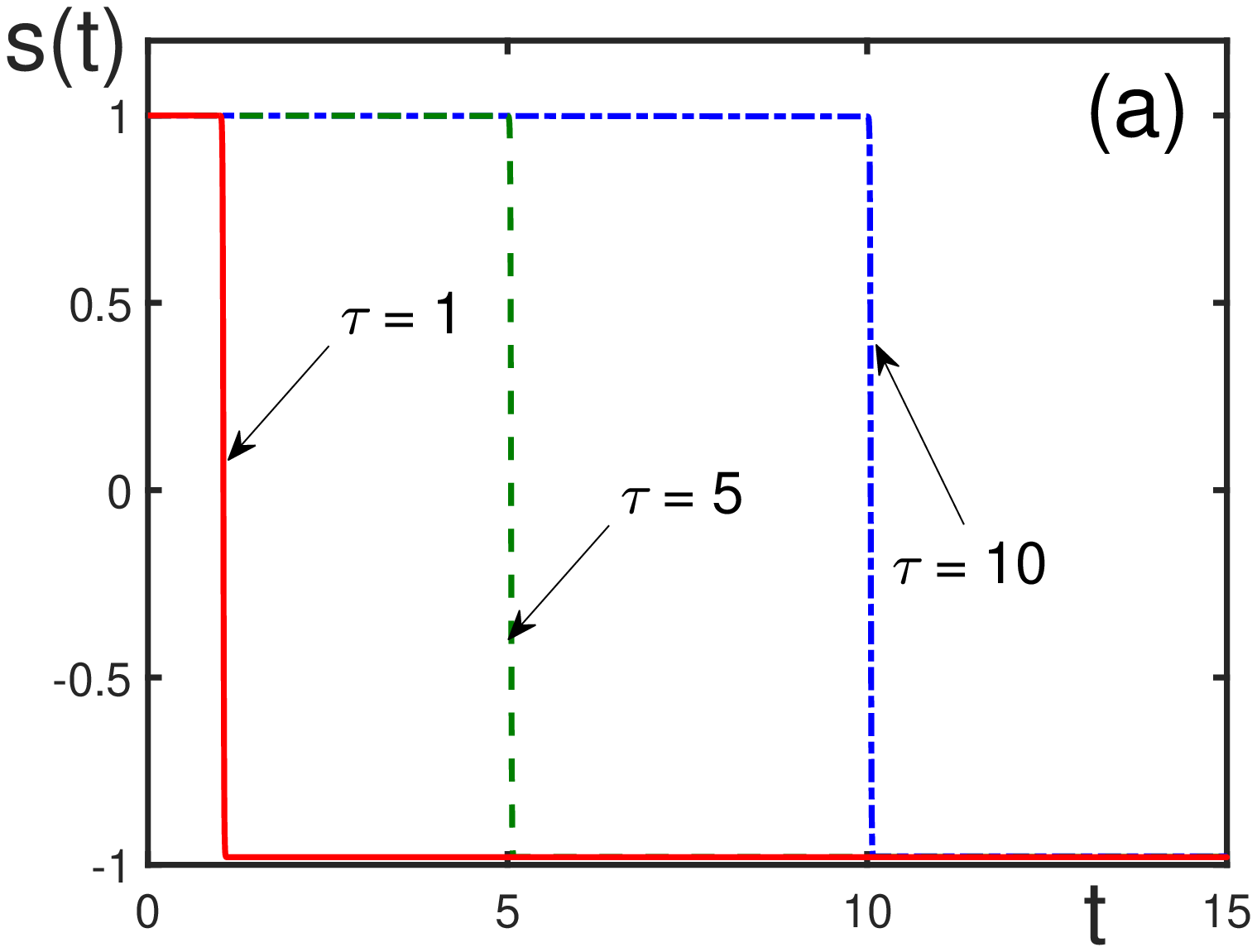} \hspace{0.5cm}
\includegraphics[width=7.0cm]{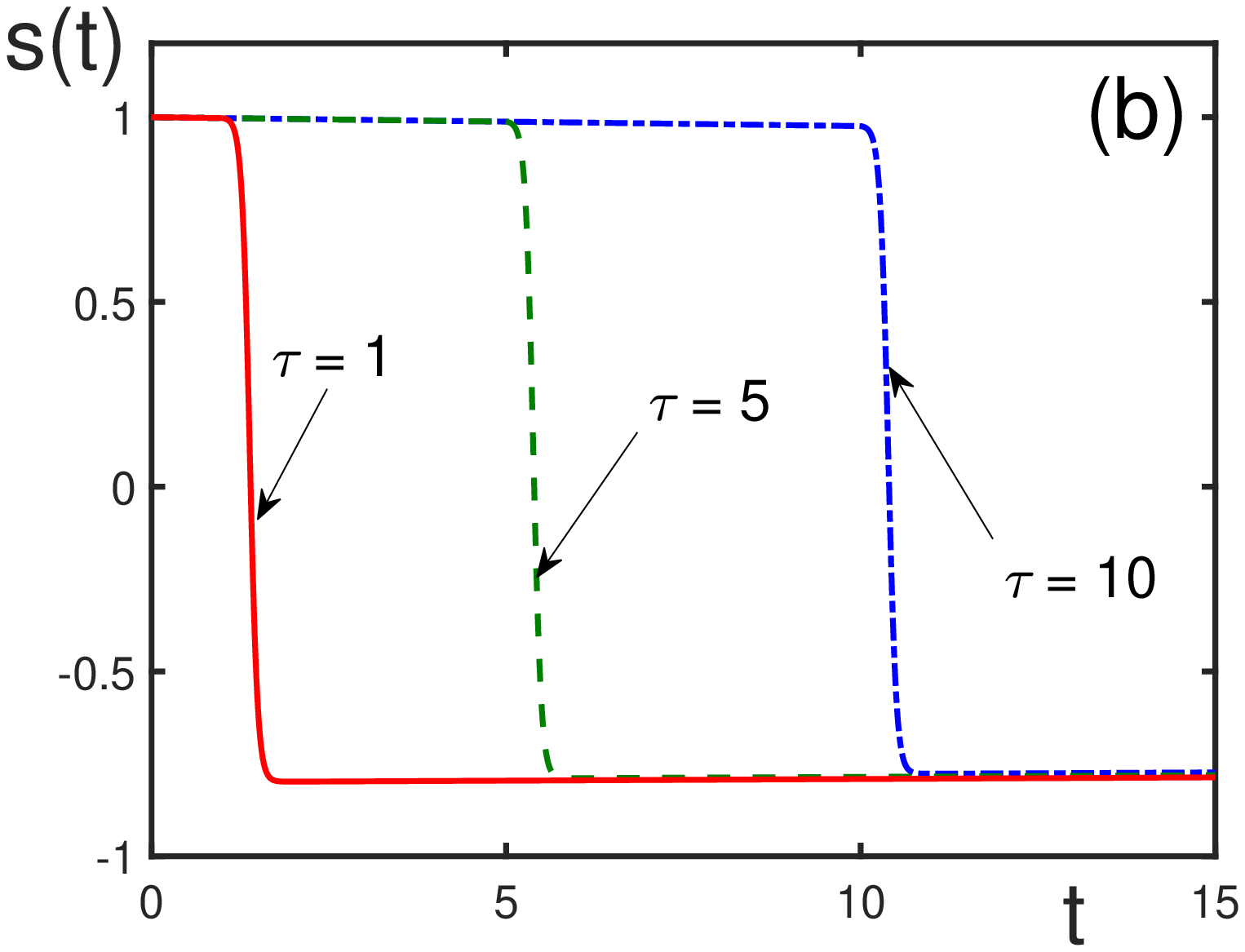}  } }
\caption{\small Average spin polarization $s(t)$ of an assembly of magnetic 
nanomolecules or nanoclusters for the anisotropy parameter $A=5$ and $\gm=10$. 
The dynamic resonance tuning starts at different delay times: $\tau=1$ (solid 
line), $\tau=5$ (dashed line), and $\tau=10$ (dash-dotted line). Resonance 
frequencies are: (a) $\om=\om_0=1000$; (b) $\om=\om_0=100$. 
}
\label{fig:Fig.5}
\end{figure}

\section{Discussion and conclusion}

We have suggested a method allowing, by using magnetic nanosystems, such as 
magnetic nanomolecules and nanoclusters, to combine two features that are crucially 
important for the functioning of memory devices, the possibility of keeping for 
long times a frozen magnetization that can be reversed at a required time. The 
method is based on the following technical stratagems. First, it is straightforward 
to use the property of magnetic nanomolecules and nanoclusters to keep, below the 
blocking temperature, the direction of magnetization frozen in a metastable state. 
Second, the sample is inserted into a magnetic coil of an electric circuit creating 
a magnetic feedback field. Third, at a required time, a varying magnetic field is 
imposed, varying in such a way that to dynamically support a resonance between the 
electric circuit and the varying Zeeman frequency of the sample. Recall that the 
effective Zeeman frequency changes in time because of the interaction between the 
moving sample spin and the magnetic anisotropy field. Due to this dynamic resonance, 
there develops a strong coupling between the sample and the electric circuit, that 
is between the sample spin and the resonator feedback field, which realizes an ultrafast 
spin reversal. This procedure can be implemented for single nanomolecules or nanoclusters 
as well as for their assemblies. The process is illustrated by numerically solving 
the spin equations of motion. 

In order to grasp the typical values of the empirical parameters, let us adduce 
several examples. Thus the typical parameters of Co, Fe, and Ni nanoclusters are 
as follows. A single cluster, of volume around $V \sim 10^{-20}$ cm$^3$, can contain 
about $N\sim 10^3 - 10^4$ atoms, so that the total cluster spin can be 
$S\sim 10^3-10^4$. The blocking temperature is $T_B\sim (10-100)$ K. With the 
magnetic field $B_0\sim 1$ T, the Zeeman frequency is $\om_0\sim 10^{11}$ s$^{-1}$. 
The feedback rate is of order $\gm_0\sim (10^{10} - 10^{11})$ s$^{-1}$. The typical 
anisotropy parameters are $D/\hbar\sim 10^7$ s$^{-1}$ and $E/\hbar\sim 10^6$ s$^{-1}$ 
or $\om_D\sim 10^{10} - 10^{11}$ s$^{-1}$ and $\om_E \sim 10^9 - 10^{10}$ s$^{-1}$. 
Hence the dimensionless anisotropy parameter can be $A\sim 0.1 - 1$. 

The magnetic nanomolecule, named Fe$_8$, possesses the spin $S=10$, blocking 
temperature $T_B\approx 1$ K, the molecule volume $V\sim 10^{-20}$ cm$^3$, the 
anisotropy parameters $D/k_B=0.27.5$ K and $E/k_B=0.046$ K, or 
$D/\hbar\sim 4\times 10^{10}$ s$^{-1}$ and $E/\hbar \sim 10^{10}$ s$^{-1}$. 
Thus, the anisotropy frequencies are $\om_D\sim 4\times 10^{11}$ s$^{-1}$ and 
$\om_E\sim 10^{11}$ s$^{-1}$. Then the dimensionless anisotropy parameter is 
$A\sim 1 - 4$.

The magnetic nanomolecule Mn$_{12}$ also has the spin $S = 10$ and the blocking 
temperature $3.3$ K. The spin polarization can be kept frozen for very long times 
depending on temperature and defined by the Arrhenius law. For example, at $T = 3$ K, 
the spin is frozen for one hour and at $2$ K, for $2$ months. The molecule radius is 
around $10^{-7}$ cm and the volume, $V \sim 10^{-20}$ cm$^3$. The Zeeman frequency, 
for $B_0 = 1$ T, is $\omega_0 \sim 10^{11}$ s$^{-1}$. The feedback rate is 
$\gamma_0 \sim 10^8$ s$^{-1}$. The anisotropy parameter $D/k_B \approx 0.6$ K and 
$D/\hbar \sim 10^{11}$ s$^{-1}$, while the value of $E$ is negligible. The anisotropy 
frequency is $\omega_D \sim 10^{12}$ s$^{-1}$. Therefore the dimensionless anisotropy 
parameter is $A \sim 10$.

These values of the parameters have been kept in mind in our numerical calculations.
The reversal time is $t_{rev} \approx \gamma/ (\gamma_0 \omega_0 s_0)$. For 
$\gamma \approx \gamma_0$ and $\omega \sim 10^{11}$ s$^{-1}$, the reversal time is
of order $t_{rev} \sim 10^{-11}$ s.

\end{document}